The Detection of Saccadic Eye Movements and Per-Eye Comparisons
using Virtual Reality Eye Tracking Devices

Final Progress Report

Teran Bukenberger (Student #: 251196551)

Thesis – CS4490Z

Department of Computer Science

Western University

February 27th, 2023

Project Supervisor: Dr. Brent Davis, Dept. of Computer Science

Course Instructor: Laura Reid, Dept. of Computer Science

# 1. ABSTRACT

Eye tracking has been found to be useful in various tasks including diagnostic and screening tools. However, traditional eye trackers had a complicated setup and operated at a higher frequency to measure eye movements. The use of more commonly available eye trackers such as those in head-mounted virtual reality (VR) headsets greatly expands the utility of these eye trackers for research and analytical purposes. In this study, the research question is focused on detecting saccades, which is a common task when analyzing eye tracking data, but it is not well-established for VR headset-mounted eye trackers. The aim is to determine how accurately saccadic eye movements can be detected using an eye tracker that operates at 60 or 90Hz. The study involves VR eye tracking technology and neuroscience with respect to saccadic eye movements. The goal is to build prototype software implemented using VR eye tracking technology to detect saccadic eye movements, and per-eye differences in an individual. It is anticipated that the software will be able to accurately detect when saccades occur and analyze the differences in saccadic eye movements per-eye. The field of research surrounding VR eye tracking software is still developing rapidly, specifically its applications to neuroscience. Since previous methods of eye tracking involved specialized equipment, using commercially and consumer available VR eye tracking technology to assist in the detection of saccades and per-eye differences would be novel. This project will impact the field of neuroscience by providing a tool that can be used to detect saccadic eye movements and neurological and neurodegenerative disorders. However, this project is limited by the short time frame and that the eye tracker used in this study operates at a maximum frequency of 90Hz.

## 2. INTRODUCTION

Eye tracking is an essential tool for understanding human behavior and cognitive processes. The ability to capture eye movements and analyze them can provide insights into attention, decision-making, and even cognitive load. Eye tracking technology has advanced considerably in recent years, particularly with the development of VR headsets. However, despite the progress made, there remain challenges in accurately detecting and analyzing eye movements using commercial VR headsets. This study aimed to detect saccades, a rapid eye movement that occurs during visual exploration, using a commercial VR headset and compare differences between each eye when a saccade occurs.

To achieve these objectives, a custom Unity application was developed, which employed the Tobii Ocumen [1] Software Development Kit (SDK) to capture eye tracking data. The study informally enrolled three participants, one of whom had amblyopia while the remaining two had no known visual impairments. Exploratory data analysis was conducted to identify patterns and prepare for data cleaning. The distribution of saccadic and non-saccadic eye movements was imbalanced, and Principal Component Analysis (PCA) was used for dimensionality reduction. The mean distance between individual eye gaze ray direction vectors was measured for each participant, and the dataset was partitioned into training and test sets. Finally, Support Vector Machine (SVM) [2] with a Radial Basis Function (RBF) kernel [3] and 4-fold cross-validation [4] were used to derive a decision boundary in a 2-dimensional plot and determine the best-fit parameters for SVM.

The results of this study have important implications for understanding the accuracy and limitations of eye tracking with commercial VR headsets. Additionally, the findings may contribute to the development of more effective methods for analyzing eye movements and detecting saccades in real-world settings.

## 3. BACKGROUND AND RELATED WORK

Human-Computer Interaction (HCI) applications and devices specifically for eye tracking have become widespread as new technology becomes more commercially available. VR headsets with built-in eye tracking capabilities are being utilized to detect saccadic eye movements [5] as well as gaze recognition for a better understanding of cognitive and visual processing [6]. This new technology has also been used to treat a rare visual disorder known as amblyopia which develops during early childhood and typically affects vision in one eye [7]. The study *Eye-Tracking Aided VR System for Amblyopic Pediatric Treatment Difficulty Adjustment* [8] applied gaze recognition to treating amblyopia to determine the training progress during amblyopic visual therapy and aid the deviating eye. In order to determine the type of eye movement that has occurred – in the case of saccades or smooth pursuit movements – rapid movements which change the point of fixation must be differentiated from slower, more controlled movements [9]. In recent studies, saccadic eye movements have been analyzed to identify neurological disorders such as Attention-Deficit Hyperactivity Disorder (ADHD); a disorder which is characterized by inappropriate levels of inattention and/or hyperactivity and impulsivity that can continue throughout life [10]. A study which analyzed eye movement behaviour using the Tobii Pico Neo 2 VR headset with a binocular eye tracking sampling rate of 60/90Hz [11] was effective for accurately detecting saccades, and was able to reveal ADHD in children [12].

## 4. RESEARCH OBJECTIVES

**O1:** Detect when a saccade occurs using a commercial VR headset.
**O2:** Compare the differences between each eye when a saccade occurs.

### 4.1. Significance

This research project is significant as it aims to detect saccades using commercially available VR headsets. This novel approach has the potential to expand the use of eye trackers for research and analytical purposes beyond traditional laboratory settings. The study's outcomes will contribute to the development of more effective methods for analyzing eye movements and detecting saccades in real-world situations.

The project's impact on the field of neuroscience is noteworthy, as it provides a tool that can be used to detect saccadic eye movements and neurological and neurodegenerative disorders. Furthermore, the study's findings may have significant implications for understanding the accuracy and limitations of eye tracking with commercial VR headsets.

## 5. METHODOLOGY

### 5.1. Data

In order to obtain eye tracking data, a custom Unity application was developed, which employed the Tobii Ocumen [1] SDK to capture eye tracking data onto the VR headset as a JSON file. The SDK has a complex data structure for storing eye tracking data that comprises numerous nested objects. Therefore, to simplify access to any field, the data structure was flattened. The VR headset has a refresh rate of 90Hz [11], generating 90 objects per second during the recording process. The study informally enrolled three participants, one of whom had amblyopia, while the other two had no known visual impairments. Each participant completed sessions of approximately 36 seconds, during which saccades were performed intentionally in various directions. To identify the time intervals in which saccades occurred, the respective participants manually labeled the datasets.

### 5.2. Exploratory Data Analysis

To identify patterns and prepare for data cleaning, an exploratory data analysis was conducted on the original dataset. The distribution of saccadic and non-saccadic eye movements was plotted in a pie chart to determine the balance of the data. It was found that 83.4% of the data belonged to the "not saccade" category, while 16.6% belonged to the "saccade" category. The imbalanced classes may cause the model to classify data into the majority category during training, reducing the model's generalizability. To address the high dimensionality of the original dataset, PCA was used for dimensionality reduction. PCA was chosen to visualize the data in 2 dimensions and identify the components with the most variability. As shown in Fig. 1, the first 4 components accounted for most of the variance. Upon calculating the sum of the variance up to the $4^{th}$ component, we get 99.69% of the total variance.

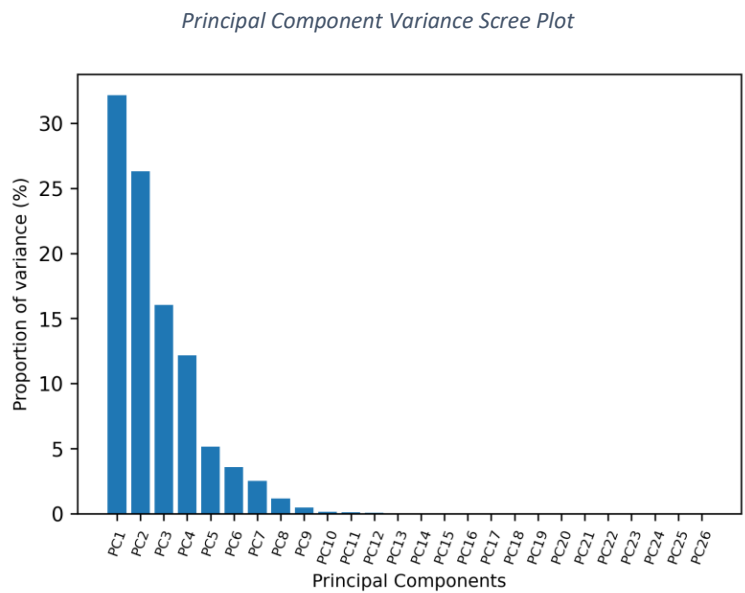

*Principal Component Variance Scree Plot*

*Fig. 1. Scree Plot of the principal components derived from eye tracking features. These features are shown in Table 1 (Appendix) and PCA was performed to see how variation was distributed across the different variables.*

### 5.3. Data Preprocessing

Before performing exploratory data analysis, it is necessary to first clean the dataset to improve accuracy.

#### 5.3.1. Invalid Gaze Ray Removal

The eye tracking data structure contains three fields pertaining to eye gaze ray direction vectors, namely *GazeRayDirectionIsValid*, *RightEyeGazeRayDirectionIsValid*, and *LeftEyeGazeRayDirectionIsValid*, which specify the validity of the gaze ray data. However, the Tobii Ocumen [1] SDK provides an ambiguous definition of what constitutes invalid gaze ray direction vectors. To ensure data integrity, rows with invalid gaze rays were excluded from the dataset.

#### 5.3.2. Noise Removal

Eye tracker latency can introduce inaccuracies in the eye tracking data, particularly in the gaze ray direction vectors of individual eyes. As a result, the dataset contains a substantial amount of noise when comparing the difference between gaze ray direction vectors per-eye. D'Agostino's $K^2$ test was performed on the dataset to see if it has a Gaussian distribution where the $p$-value was 0.0, showing that the dataset does not have a Gaussian distribution. As the dataset does not conform to a Gaussian distribution, interquartile range was used to identify extreme outliers when calculating the difference between vectors. The corresponding rows were subsequently filtered from the dataset when calculating the average distance.

### 5.4. Average Gaze Ray Direction Difference

The mean distance between individual eye gaze ray direction vectors was measured for each participant. This assessment was conducted to determine differences in eye movement patterns between participants with and without visual impairments. The left and right gaze ray direction vectors were analyzed to identify any variations and assess how they differed among participants. The minimum, maximum, and average differences were recorded and analyzed to identify any potential differences in eye movements between the participants.

### 5.5. Train and Test Data Separation

To enhance the model's accuracy, the original dataset was partitioned into training and test sets, constituting 75% and 25% of the dataset, respectively. The randomization of the data sections ensured that the training and test datasets were representative of the overall dataset.

### 5.6. Support Vector Machine Classification

SVMs [2] are a commonly used machine learning algorithm in classification and regression tasks. In this study, SVM was used with a RBF kernel [3]. The RBF kernel allows SVM to learn nonlinear decision boundaries, which can better fit complex data. SVM has several hyperparameters that need to be optimized to achieve the best performance. To find the best hyperparameters, a grid search [3] approach was employed.

5.7. **Cross-Validation**

Cross-validation [4] is a statistical method used to evaluate the performance of a machine learning model by testing it on an independent dataset. 4-fold cross-validation was used to validate the performance of the SVM model. The first step in cross-validation is to split the dataset into a training set and a testing set (Section 5.5). In 4-fold cross-validation, the dataset is divided into four equal parts, with three parts used for training the model, and one part used for testing the model. This process is repeated four times, with each of the four parts used for testing, and the remaining three parts used for training. PCA was also utilized to reduce the dimensionality of the dataset and derive a decision boundary to be later used in a 2-dimensional contour plot. The cross-validation results were used to determine the best-fit parameters for the SVM model with a RBF kernel.

## 6. RESULTS

The purpose of this study was to examine inter-participant differences in per-eye movements and to identify instances of saccades. Eye tracking data was obtained from a sample of three participants, of whom one had amblyopia while the remaining two had no known visual impairments. Participants underwent sessions lasting approximately 36 seconds, during which they intentionally performed saccades in different directions. To identify instances of saccade occurrence, datasets were labeled.

Exploratory data analysis was conducted to identify patterns and prepare for data cleaning. The distribution of saccadic and non-saccadic eye movements was 16.6% and 83.4% respectively. To address the 25 variables in the original dataset, PCA was used for dimensionality reduction. The first 4 components accounted for 99.69% of the total variance. Figure 2 shows the eye tracking data in a scatter plot for the 4 principal components.

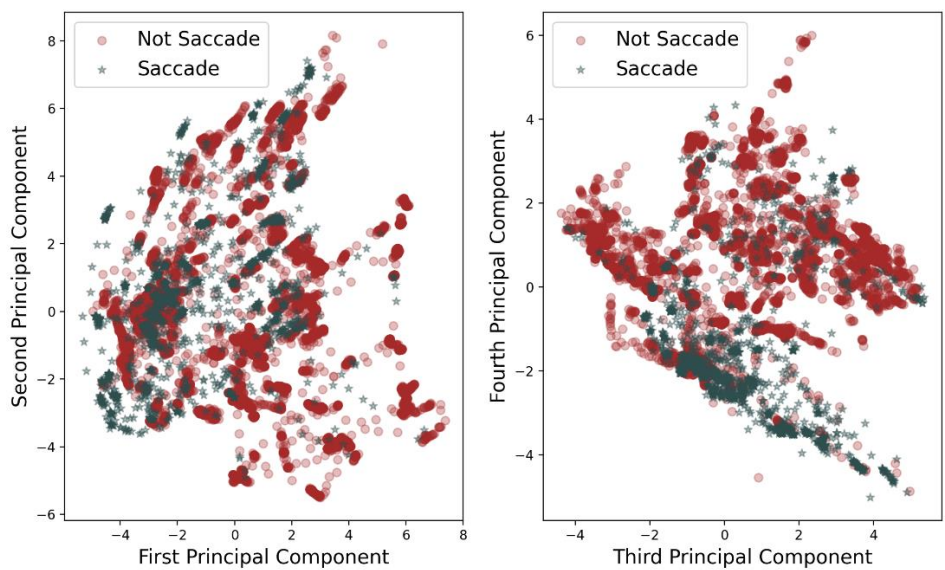

*Fig. 2. The eye tracking data in a scatter plot. Saccade and Non-Saccade variation is shown for the first 4 principal components.*

The mean distance between individual eye gaze ray direction vectors was assessed for each participant. The findings suggest that the amblyopic participant (Fig. 2a) displayed larger differences between the left and right gaze ray direction vectors compared to the other two participants, with minimum, maximum, and average differences of 0.0639, 0.1539, and 0.0003, respectively. The first participant (Fig. 2b) with no vision impairments showed relatively smaller differences, with minimum, maximum, and average differences of approximately 0.0229, 0.0523, and 0.0012, respectively. The second participant (Fig. 2c) with no vision impairments exhibited minimum, maximum, and average differences of approximately 0.0401, 0.0617, and 0.0203, respectively.

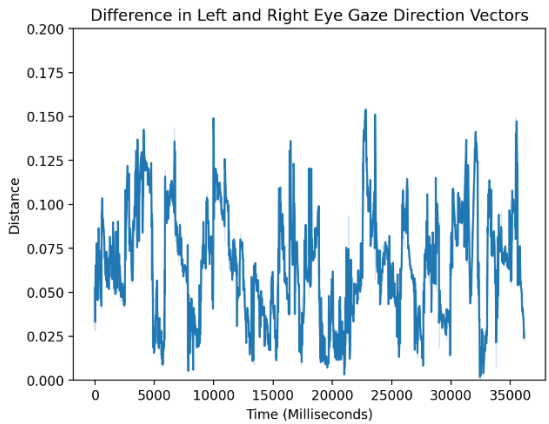

*Fig. 3a. Gaze ray direction vector difference line plot for the amblyopic participant.*

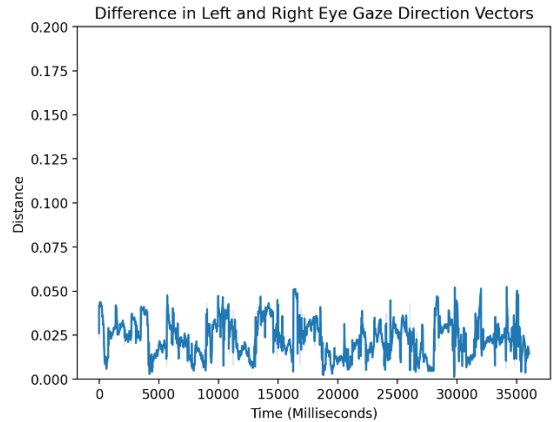

*Fig. 3b. Gaze ray direction vector difference line plot for the first participant with no vision impairment.*

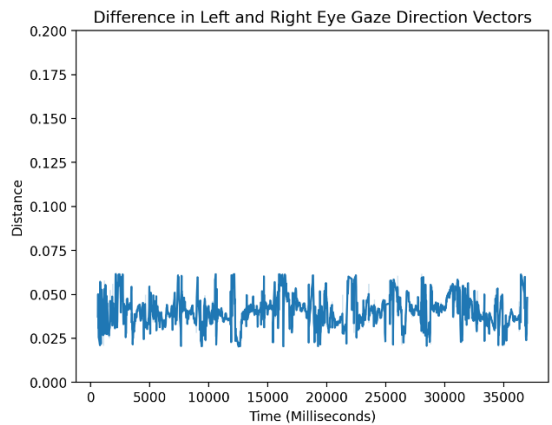

*Fig. 3c. Gaze ray direction vector difference line plot for the second participant with no vision impairment.*

To enhance the model's accuracy, the original dataset was partitioned into training and test sets, constituting 75% and 25% of the dataset, respectively. The randomization of the data sections ensured that the training and test datasets were representative of the overall dataset. SVM was used with a RBF kernel, which allows SVM to learn nonlinear decision boundaries, better fitting complex data. The hyperparameters of SVM were optimized using a grid search approach. The application of 4-fold cross-validation and PCA were employed to derive a decision boundary in a 2-dimensional contour plot (Fig. 4) and to determine the best-fit parameters for SVM with an RBF kernel. Four principal components were utilized, and after conducting 4-fold cross-validation, the SVM classifier demonstrated 93% accuracy. The resulting confusion matrix shown in Fig. 5 provided the precision score, recall score, and F1 score of 0.9337, 0.9313, and 0.9313, respectively, with scores averaged by weighted support to address label imbalance, as described in section 5.2.

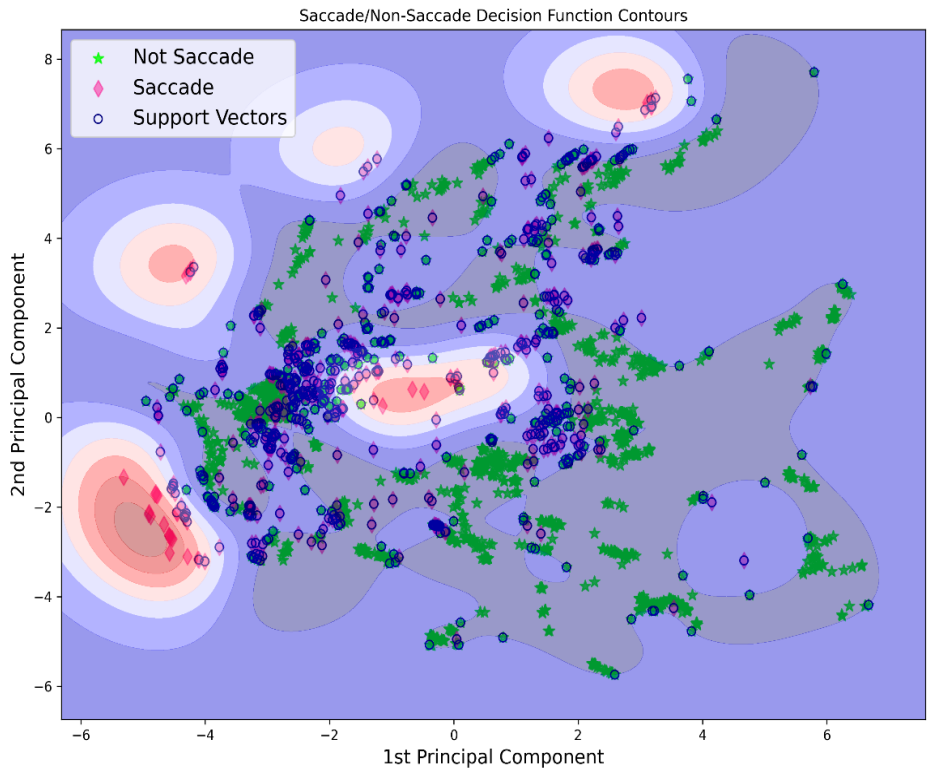

*Fig. 4. Decision Function contours with the RBF kernel is shown here along with support vectors. Best-fit parameters are obtained from 4-fold cross-validation.*

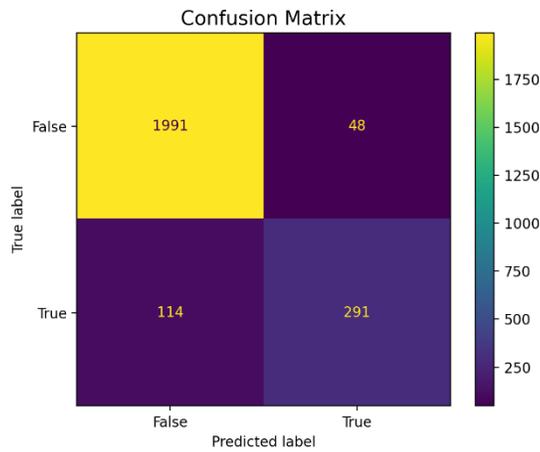

*Fig. 5. Confusion Matrix obtained using a pipeline with the SVM classifier and RBF kernel.*

## 7. DISCUSSION

This study aimed to evaluate the accuracy of an SVM model for saccade detection using 4-fold cross-validation and PCA to extract the principal components. Despite the SVM classifier claiming 93% accuracy, the visualization of the Decision Function contours (Fig. 4) did not display clear boundaries between saccades and non-saccades, indicating a lack of accuracy in saccade detection. We speculate that the inconsistent data resulting from the absence of a formal process for recording eye movements during the experiment might have contributed to this outcome. This inconsistency also led to an imbalanced dataset, as discussed in Section 5.2.

The variation plot of the first four principal components (Fig. 2) revealed that saccades and non-saccades overlapped, which could have further complicated the classification task. Our findings suggest that future studies should focus on implementing a formal process for recording eye movements to improve data consistency and address the issues of dataset imbalance.

Furthermore, we analyzed the average distance between gaze ray direction vectors per-eye, and the results displayed in Fig. 3a, 3b, and 3c revealed that the participant with amblyopia had a significantly greater minimum and maximum distance than those with no visual impairments. These findings suggest that commercial eye tracking devices could be used to detect amblyopia in future research.

## 8. CONCLUSION

This study investigated the ability of commercially available VR headsets to detect saccades and differences in saccadic eye movements per-eye. Results showed that the headset in combination with SVM can detect saccades with 93% accuracy (Fig. 5). Additionally, the study explored differences in saccadic eye movements per-eye, providing valuable insights into visual perception.

These findings have significant implications for neuroscience and human-computer interaction as VR headsets can potentially analyze eye movements and detect saccades in real-world settings. The study also provides a framework for future research and contributes to the development of more effective methods for analyzing eye movements.

Limitations of the study include a small sample size and a maximum frequency of 90Hz for the eye tracker. Future studies with larger sample sizes and higher frequency eye trackers can improve accuracy and provide more robust insights into visual perception and cognitive processes.

Overall, this study demonstrates the potential of commercial VR headsets as a tool for analyzing eye movements and detecting saccades, providing a foundation for future research in this area.

## 9.  FUTURE WORKS AND LESSONS LEARNED

**Future Work**
To improve the accuracy of saccade detection, there are several potential areas for future research. Firstly, incorporating additional features such as saccade velocity, eye movement history, and cognitive state into the predictive model could enhance its predictive power. Secondly, exploring different machine learning algorithms and techniques could lead to further improvements in saccade detection through landing point prediction. Thirdly, testing the generalizability of the model across diverse populations would provide valuable insights into its applicability in various contexts. Addressing these areas of future work would advance our understanding of saccade dynamics and improve the development of eye-tracking technologies for diverse applications.

**Lessons Learned**
Exploratory data analysis is a fundamental step in understanding the data that will be used in machine learning models. Through exploratory data analysis, valuable insights can be obtained that can guide the selection of appropriate data processing techniques. As such, it is imperative to invest time and resources in this process to ensure the best results. The study has shown that by performing exploratory data analysis, we were able to better understand the data and select suitable data processing techniques.

Secondly, inconsistencies across different datasets can pose significant challenges to the validity and reliability of research findings. To address this issue, it is crucial to formalize the process of recording data from study participants. This process should be standardized and consistently applied across all participants to ensure that the resulting datasets are comparable and can be used effectively in machine learning models. In our study, we encountered challenges with inconsistent datasets, which hindered our ability to compare and analyze data effectively. Thus, we emphasize the importance of formalizing data recording processes to ensure consistency and reliability in research findings.

## 10. ACKNOWLEDGEMENTS

Page 15 of 16                                                                                           Teran Bukenberger

# APPENDIX

### Table 1: Features of the eye tracking data set

| Feature | Description |
|---|---|
| leftGazeRayDirectionX | X component of the left eye gaze ray direction |
| leftGazeRayDirectionY | Y component of the left eye gaze ray direction |
| leftGazeRayDirectionZ | Z component of the left eye gaze ray direction |
| leftGazeRayOriginX | X component of the left eye gaze ray origin |
| leftGazeRayOriginY | Y component of the left eye gaze ray origin |
| leftGazeRayOriginZ | Z component of the left eye gaze ray origin |
| leftPupilDiameter | The diameter of the left eye pupil |
| leftPositionGuideX | X component of the normalized position of left pupil in relation to optical axis |
| leftPositionGuideY | X component of the normalized position of left pupil in relation to optical axis |
| rightGazeRayDirectionX | X component of the right eye gaze ray direction |
| rightGazeRayDirectionY | Y component of the right eye gaze ray direction |
| rightGazeRayDirectionZ | Z component of the right eye gaze ray direction |
| rightGazeRayOriginX | X component of the right eye gaze ray origin |
| rightGazeRayOriginY | Y component of the right eye gaze ray origin |
| rightGazeRayOriginZ | Z component of the right eye gaze ray origin |
| rightPupilDiameter | The diameter of the right eye pupil |
| rightPositionGuideX | X component of the normalized position of right pupil in relation to optical axis |
| rightPositionGuideY | Y component of the normalized position of right pupil in relation to optical axis |
| gazeRayDirectionX | X component of the gaze direction for both eyes |
| gazeRayDirectionY | Y component of the gaze direction for both eyes |
| gazeRayDirectionZ | Z component of the gaze direction for both eyes |
| gazeRayOriginX | X component of the gaze origin for both eyes |
| gazeRayOriginY | Y component of the gaze origin for both eyes |
| gazeRayOriginZ | Z component of the gaze origin for both eyes |
| convergenceDistance | The distance along the combined gaze ray where the eyes converge, measured in meters |